\documentclass[twocolumn,showpacs,prb]{revtex4} 

\begin{document}

\title{Efficient algorithm for current spectral density calculation \\
in single-electron tunneling and hopping }
\author{Viktor A. Sverdlov}
\affiliation{Institute for Microelectronics, TU Vienna, 
       Gusshausstrasse 27-29, A-1040 Vienna, Austria} 
\author{Yusuf A. Kinkhabwala}
\affiliation{Department of Physics and Astronomy, Stony Brook
University, Stony Brook, NY
11794-3800
}
\author{Alexander N. Korotkov}
\affiliation{Department of Electrical Engineering, University of
California, Riverside, CA 92521 
}
\date{\today}

\begin{abstract}
This write-up describes an efficient numerical method for 
the Monte Carlo calculation 
of the spectral density of current in the multi-junction single-electron
devices and hopping structures. In future we plan to expand this write-up
into a full-size paper.

\end{abstract}

\maketitle



	In this paper we describe an algorithm for the 
Monte Carlo calculation of the spectral density $S_I(\omega )$ of tunneling 
current in multi-junction single-electron devices.\cite{Av-Likh} 
The same algorithm is applicable to calculation of  
the noise at hopping \cite{hopping} because of the problem similarity. 
This algorithm has been used in several of our earlier papers; 
\cite{Kor-Likh-00,Sverdlov-01,S-K-K-L-01,KSLK-03,KSL-04} 
however, it has not yet been described explicitly 
(except for revised versions of unpublished paper \onlinecite{KSLK-03}). 

	\vspace{0.3cm}

The first spectral calculations of the electron transport in
single-electron devices using the Monte Carlo technique have been
performed in Refs.~\onlinecite{Bakhvalovetal1989}
and~\onlinecite{Amman-89}; in these papers the spectral density
has been calculated as a Fourier transform of the correlation 
function. However, this method is rather slow in the case when the
current $I\left( t\right)$ is a sequence of $\delta$-functions,
corresponding to tunneling events:
\begin{equation}
I\left( t \right) = \sum_n q_n \delta \left( t-t_n \right),
\end{equation}
where $t_n$ is a (random) time of the $n$-th tunneling event
and $q_n$ is the corresponding charge transfer. (The sequence
$\left\{ q_n\right\}$ is also random and reflects the path in the
space of charge configurations.)

A significantly faster ``standard'' algorithm \cite{Kor-94}
(embedded, for example, into the simulation package MOSES
\cite{MOSES}) is based on the definition of the spectral density
$S_I\left( \omega \right)$ of the current $I\left( t \right)$ via
the square of the Fourier transform $\left\vert I\left( \omega
\right)\right\vert^2$.  More specifically,
for the rectangular time window (natural in simulations)
there is a relation
\begin{eqnarray}
&& \frac{2}{T}\, \left\langle \left\vert \int_{t_0}^{t_0+T}
I\left(t \right) \exp \left(i \omega t \right)\, dt \right\vert^2
\right\rangle
\nonumber \\
&& = \int_{-\infty}^{+\infty} S_I\left( \omega +\Omega \right)\,
\frac{1-\cos
\left( \Omega T\right) } {\pi T\Omega^2} \, d\Omega
 \label{S(w)}
\end{eqnarray}
(here $\langle ... \rangle$ denotes ensemble averaging and $i$ 
is the imaginary unit), whose right hand side tends 
to $S_I\left( \omega \right)$ in the limit $T\rightarrow \infty$.
Therefore,
\begin{equation}
\tilde{S}_I \left( \omega \right) \equiv \frac{2}{T} \left\langle
\left\vert \sum_n q_n \exp \left(i \omega t_n \right)
\right\vert^2 \right\rangle \label{Stilde}
\end{equation}
is a good approximation for the true spectral density $S_I\left(
\omega \right)$ even for a finite, but large enough time interval
$T$. (Summation in Eq.\ (\ref{Stilde}) is over the tunneling
events within the interval $t_0< t_n <t_0+T$). In the standard
method \cite{Kor-94,MOSES} the ensemble averaging in Eq.\
(\ref{Stilde}) is replaced by averaging over $K$ sequential time
segments (each of duration $T$) of the Monte Carlo realization, so
that $t_{0}$ becomes $jT$, where $j=1$, 2, $... \, K$. It is
natural to calculate simultaneously the spectral density for a set
of frequency points (the set of harmonics of a certain low
frequency is most convenient), and it is useful to choose $\omega
/2\pi$ equal to integer multiples of $T^{-1}$ to avoid
``poisoning" of the right hand side of Eq.\ (\ref{S(w)}) by the
$\delta$-function contribution from $S_I\left( 0 \right)$ due to
dc current $\overline I$. (Other ways of subtracting the effect of
$\overline I$ are also possible.)

A major disadvantage of this standard method is that the relative
accuracy of the spectral density calculation cannot be better than
approximately $K^{-1/2}$, because the right hand side of Eq.\
(\ref{Stilde}) before averaging over $K$ segments has the rms
fluctuation comparable to the mean value. It is easy to increase
$K$ (without increasing the total simulation time) by decreasing
$T$; however, besides increasing the smoothing of $S_I\left(
\omega \right)$ [which is $\Delta \omega \sim T^{-1}$ -- see Eq.\
(\ref{S(w)})], this may lead to incorrect results when $T$ becomes
comparable or less than the longest correlation time of the
simulated process, and therefore the $T$-segments are no longer
statistically independent. Since the correlation time is not known
in advance (it may be estimated as the lowest frequency at which
the spectral density levels off), the choice of $T$ is not a
trivial task and requires some intuition that complicates the use
of the standard method.

Here we describe the advanced algorithm of spectral density
calculation which eliminates this problem and also makes
calculation significantly faster (for the same accuracy of the
result). The method is somewhat similar to the ``reduced'' method
for dc current calculation \cite{Kor-94} and basically treats the
randomness of tunneling times $t_n$ analytically, while the path
in the charge configuration space is still simulated
\cite{Bakhvalovetal1989} by the Monte Carlo technique.

Let us consider a $T$-long realization of the process assuming for
simplicity $t_0=0$, so that $t_n=\sum_{k=1}^n \tau_k$ where
$\tau_k$ is the time between the adjacent tunneling events, i.e.
time spent in a particular charge state. In the case when the
system parameters (external voltage, etc.) do not change with
time, the random time $\tau_k$ has the Poisson distribution with
the average value $\left\langle \tau_k \right\rangle =
1/\Gamma_{k,\Sigma}$, where $\Gamma_{k,\Sigma}$ is the sum of all
tunneling rates for the corresponding charge state. The quantity
$s \equiv \left\vert \sum_n q_n \exp \left( i\omega t_n \right)
\right\vert^2$, which is related to the spectral density via Eq.\
(\ref{Stilde}), may be easily expressed as
\begin {eqnarray}
s &=& \sum_{n,m} q_n q_m \exp \left[ i\omega \left(
\sum_{k=1}^n\tau_k - \sum_{k=1}^m \tau_k \right) \right]
\nonumber \\
 &=& \sum_n q_n^2 + 2 \, \mbox{Re} \sum_{n>m} q_n q_m \exp \left[
i\omega \sum_{k=m+1}^n \tau_k \right].
\hspace{0.5cm} \label{s-1}
\end{eqnarray}

For the ensemble averaging of $s$ let us first average Eq.\
(\ref{s-1}) over random $\tau_k$, leaving averaging over paths in
charge space for later. Using the mutual independence of $\tau_k$
fluctuations, we can average each exponent independently:
\begin{equation} \left\langle e^{i\omega \tau_k} \right\rangle = \int_0^\infty
\frac{e^{-\tau /\left\langle\tau_k \right\rangle }}{\left\langle
\tau_k \right\rangle} e^{i\omega \tau} d\tau = \frac{1}{1-i\omega
\left\langle \tau_k \right\rangle},
\end{equation}
thus obtaining the expression
\begin{equation}
\left\langle s \right\rangle = \sum_n q_n^2 +2 \, \mbox{Re} \left(
\sum_{n>m} q_n q_m \prod_{k=m+1}^n \frac{1}{1- i\omega\left\langle
\tau_k \right\rangle } \right) . \label{s-aver}
\end{equation}

This expression can be calculated iteratively introducing
complex variables
\begin{eqnarray}
A_p &\equiv& \sum_{n=1}^p q_n^2 + 2\sum_{n>m}^p q_n q_m \prod
_{k=m+1}^n \frac{1}{1- i\omega \langle \tau_k\rangle},
\hspace{0.5cm} \\
B_p &\equiv& \sum_{m=1}^p q_m \prod_{k=m+1}^p \frac{1} {1- i\omega
\left\langle \tau_k\right\rangle} ,
\end{eqnarray}
that satisfy recurrent equations
\begin{eqnarray}
A_{p+1} &=& A_p +q_{p+1}^2 + 2q_{p+1} B_p \, \frac{1}{1- i\omega
\left\langle
\tau_{p+1}\right\rangle } , \hspace{0.5cm}
\label{recur1} \\
B_{p+1} &=& q_{p+1} +B_p \, \frac{1}{1- i\omega \left\langle
\tau_{p+1} \right\rangle } , \label{recur2}
\end{eqnarray}
with initial condition $A_0=B_0=0$, while $\left\langle
s\right\rangle =\mbox{Re} A_{p}$ at the end of realization.

It is important to notice that $\mbox{Re} A_p$ accumulates with
the length of realization (in contrast to $s$ before averaging,
which is a strongly fluctuating variable), so that $\left(
2/\left\langle t_p\right\rangle \right)\,\mbox{Re} A_p$ (where
$\left\langle t_p\right\rangle = \sum_k^p \left\langle
\tau_k\right\rangle $) tends to some limit at $p\rightarrow
\infty$. This is the reason why, in contrast to the standard
method, the numerical averaging over many $T$-segments is not
necessary now, and the ensemble averaging of the segments over
different realizations can be replaced by the natural ``time''
averaging over the length of a realization. This eliminates the
problem of choosing $T$, discussed above, and now $T$ can be
treated as a running variable $T_p=\left\langle t_p\right\rangle$
during the whole simulation run. Similarly, $s$ can also be
treated as a running variable $s_p$.
 (Strictly speaking, averaging
over $\tau_k$ in the segments with a fixed time $T$ and/or a fixed
charge path is different; however, the difference vanishes at large $T$).

Thus, the basic algorithm is the following. The Monte Carlo
technique is used to simulate one long realization of the random
path in the configuration (charge) space, while the time is
treated deterministically as $\sum_k \left\langle
\tau_k\right\rangle$; the variables $A_p$ and $B_p$ are updated
after each tunneling event using Eqs.\
(\ref{recur1})--(\ref{recur2}), and the current spectral density
$S_I\left(\omega \right)$ is calculated as
    \begin{equation}
S_I\left(\omega \right) \approx
\frac{2}{\left\langle t_p \right\rangle }\, \mbox{Re} A_p.
    \end{equation}
Even though breaking the simulation into segments is not needed in the new method,
the calculation and comparison
of partial results for $S_I\left(\omega \right)$ on some time
segments is useful for run-time estimates of the calculation accuracy.

Actually, this basic algorithm still requires several improvements
to become faster than the standard method, especially at low
frequencies. First, the accuracy can be significantly improved by
explicitly calculating the spectral density for the function
$I\left( t\right)-\overline{I}$ instead of $I\left( t\right)$.
(The average current $\overline{I}$ can be calculated as $\sum_k
q_k/\sum_k \left\langle \tau_k\right\rangle $, which is the same
as in the reduced method.\cite{Kor-94,MOSES}) For this purpose
the definition of quantity $s_p$ should be modified to $s_p=
\left\vert \left[\sum_n^p q_n \exp \left( i\omega t_n
\right)\right] - \overline{I} \, \left[\exp \left(i\omega
t_p\right)-1\right]/i\omega \right\vert^2 = \left\vert \sum_n^p
\exp \left( i\omega t_n \right) \left[q_n-\overline{I}
\left(1-\exp\left(-i\omega\tau_n\right)\right)/i\omega
\right]\right\vert^2$. From this point, the derivation is
similar to that discussed above, though is now significantly
lengthier. The final result is that the only change in the
algorithm is a different set of recurrent equations replacing
Eqs.\ (\ref{recur1})--(\ref{recur2}):
\begin{eqnarray}
A_{p+1} &=& A_p +q_{p+1}^2 -2\overline{I} \left\langle
\tau_{p+1}\right\rangle \, \frac{q_{p+1}-\overline{I} \langle
\tau_{p+1}\rangle } {1+\left( \omega \left\langle
\tau_{p+1}\right\rangle \right)^2} \hspace{1.2cm}
\nonumber \\
&& + 2\frac{q_{p+1}-\overline{I} \left\langle
\tau_{p+1}\right\rangle } {1-i\omega \left\langle
\tau_{p+1}\right\rangle }B_p, \,
\label{Anew} \\
B_{p+1} &=& q_{p+1} - \frac{\overline{I} \left\langle
\tau_{p+1}\right\rangle } {1-i\omega \left\langle
\tau_{p+1}\right\rangle } + B_p \frac{1}{1-i\omega  \left\langle
\tau_{p+1}\right\rangle }.
\label{Bnew}
\end{eqnarray}
(The initial conditions are still $A_0=B_0=0$).

However, this improvement still does not solve the problem of
relatively poor convergence of the algorithm, especially at low
frequencies. The origin of the problem is hinted at by Eq.\
(\ref{S(w)}). Since we eliminated the $T$-segmentation used in the
standard method, and now $T$ is much longer (the whole simulation
period), we are calculating $S_I(\omega )$ with a much smaller
degree of spectral smoothing. The price for a better spectral
resolution $\Delta \omega$ is the longer simulation time for the
same accuracy. Therefore, to improve convergence, we have to
re-introduce some time constant $T_0$
that would define the spectral smoothing $\Delta \omega
\sim 1/T_0$. In principle, there are many ways to do this. For
example, we can periodically (with period $T_0$) set to zero the
value of $B_p$ (in this case the algorithm becomes somewhat
similar to the standard method). Alternatively, we can introduce a
gradual cutoff of $B_p$, for example, multiplying the last term in
Eq.\ (\ref{Bnew}) by $\exp (-\left\langle \tau_{p+1}\right\rangle
/T_0)$, and so on.

We have used the following way of introducing $T_0$, which is the best
among those we had tried. For simplicity, let us consider first the
algorithm without subtraction of $\overline{I}$, and average Eq.\
(\ref{s-aver}) over frequency (from $\omega =-\infty$ to $\omega
=\infty$) with the Lorentzian weight factor $\left(T_0/\pi
\right)/\left[1+\left(\omega -\tilde\omega\right)^2 T_0^2\right]$.
The integral can be easily calculated using the residue theorem
since all the poles of Eq.\ (\ref{s-aver}) are in the lower half
of the complex plane; therefore, closing the integration contour
in the upper half-plane, we will have only one pole at $\omega =
\tilde{\omega}+ i/T_0$. As a result, the only change in  Eq.\
(\ref{s-aver}) after integration is that $\omega$ is replaced by
$\omega + i/T_0$ (more correctly, by $\tilde\omega + i/T_0$, but
for simplicity we change the notation from $\tilde\omega$ back to
$\omega$). Therefore, the Lorentzian averaging over frequency in
our algorithm exactly corresponds to replacing $\omega$ with
$\omega + i/T_0$ in the iteration equations
(\ref{recur1})--(\ref{recur2}).

For the algorithm with $\overline{I}$ subtraction, the Lorentzian
averaging is a little more difficult, because of the extra poles
in the equation for $\left\langle s\right\rangle$ at $\omega =
i/\left\langle \tau_k\right\rangle$ (upper half-plane) and at
$\omega=0$. However, as seen from Eqs.\
(\ref{Anew})--(\ref{Bnew}), the pole at $\omega=0$ is eventually
canceled, while the poles at $\omega=i/\left\langle
\tau_k\right\rangle$ remain only in the simple additive term in
Eq.\ (\ref{Anew}). Therefore, the recipe of replacing $\omega$
with $\omega+i/T_0$ still works for $B_{p+1}$, and the extra
residue of the upper-half-plane pole should be simply added to
$A_{p+1}$. As a result, Eqs.\ (\ref{Anew})--(\ref{Bnew}) are
replaced with
\begin{eqnarray}
&& \hspace{-0.3cm}
A_{p+1} = A_p+q_{p+1}^2 + 2 \frac{q_{p+1}-\overline{I}
\left\langle \tau_{p+1}\right\rangle }
{1-i\left(\omega+i/T_0\right) \left\langle \tau_{p+1}\right\rangle
}B_p
\nonumber \\
&& \hspace{-0.1cm}
- \frac{2\overline{I} \left\langle \tau_{p+1}\right\rangle
\left( q_{p+1}-\overline{I} \left\langle \tau_{p+1}\right\rangle
\right)\left( 1+\left\langle \tau_{p+1}\right\rangle /T_0 \right)}
{1+\left( \omega \left\langle \tau_{p+1}\right\rangle \right)^2 +
2 \left\langle \tau_{p+1}\right\rangle /T_0 +\left( \left\langle
\tau_{p+1}\right\rangle /T_0 \right)^2 } \, , \hspace{0.5cm}
\,\,\,\,\,\,\,  \label{Anewnew} \\
&& \hspace{-0.3cm}
 B_{p+1} = q_{p+1} - \frac{\overline{I} \left\langle
\tau_{p+1}\right\rangle } {1-i\left(\omega+i/T_0\right)
\left\langle \tau_{p+1}\right\rangle } \nonumber \\
&& \hspace{0.7cm}
+ B_p \frac{1}{1-i\left(\omega +i/T_0\right)
\left\langle \tau_{p+1}\right\rangle } \, ,
\label{Bnewnew}
\end{eqnarray}
while the rest of the algorithm does not change.

The introduction of Lorentzian smoothing greatly improves the
convergence of the algorithm. However, it gives rise to another
difficulty. The problem is that the averaging over frequency
increases the $\delta$-function contribution from
$S_I\left(0\right)$ due to average current, and the trick of the
standard method, discussed above, is impossible for Lorentzian
averaging [in contrast to Eq.\ (\ref{S(w)}), in which the
convolution function contains zeros]. Formally, our algorithm
subtracts $\overline{I}$ beforehand; however, in a real simulation
$\overline{I}$ is not known exactly (note that the
estimated value of $\overline{I}$ improves during the 
course of simulation). It can be shown that the inaccuracy $\Delta
I$ in the average current estimate used in Eqs.\
(\ref{Anewnew})--(\ref{Bnewnew}) brings to $S_I\left(\omega
\right)$ the extra contribution
\begin{equation}
\Delta S_I\left(\omega \right) = 4 T_0 \left(\Delta
I\right)^2/\left(1+\omega^2 T_0^2\right). \label{Delta-S}
\end{equation}
This contribution can be subtracted from $S_I\left(\omega \right)$
at the end of the simulation run, when a better estimate of
$\overline I$ is known and the difference from the initially used
estimate can be calculated. Actually, the value of $\overline{I}$
used in Eqs.\ (\ref{Anewnew})--(\ref{Bnewnew}) can be periodically
(sufficiently rare) updated during the simulation run; in this
case $(\Delta I)^2$ in Eq.\ (\ref{Delta-S}) can naturally be
replaced with the time-weighted value.

With these modifications, the advanced algorithm becomes
significantly faster and more convenient than the standard
algorithm. Accurate comparison of their efficiencies is not
straightforward because both methods have adjustable parameters.
($T$ in the standard method and $T_0$ in the new method both affect
the smoothing of the spectral density and the convergence speed; the
choice of too short $T$ could also lead to incorrect results.)
Crudely, the speed-up factor (the ratio of CPU times for the same
accuracy using the two methods) for our typical simulation runs is
two to three orders of magnitude.

	\vspace{0.3cm}

	The authors thank K. K. Likharev for useful discussions and
for critical reading and improvement of this text.


\end{document}